# Perturbative Dynamics of Stationary States in Nonlinear Parity-Time Symmetric Coupler

Jyoti Prasad Deka and Amarendra K. Sarma
*Department of Physics, Indian Institute of Technology Guwahati, Guwahati-781 039, Assam, India*

We investigate the nonlinear parity-time (PT) symmetric coupler from a dynamical perspective. As opposed to linear PT-coupler where the PT threshold dictates the evolutionary characteristics of optical power in the two waveguides, in a nonlinear coupler, the PT threshold governs the existence of stationary points. We have found that the stability of the ground state undergoes a phase transition when the gain/loss coefficient is increased from zero to beyond the PT threshold. Moreover, we found that instabilities in initial conditions can lead to aperiodic oscillations as well as exponential growth and decay of optical power. At the PT threshold, we observed the existence of a stable attractor under the influence of fluctuating gain/loss coefficient. Phase plane analysis has shown us the presence of a toroidal chaotic attractor. The chaotic dynamics can be controlled by a judicious choice of the waveguide parameters.

## 1. INTRODUCTION

Bender and Boettcher's pioneering work [1] on a class of non-Hermitian Hamiltonian paved the way for new developments in the foundational studies of quantum mechanics [2-6]. They showed that such Hamiltonians possess a real eigenspectra as long it respects the criteria of *PT* (Parity and Time Reversal) symmetry. In general, the Hamiltonian $H = -\frac{1}{2}\frac{d^2}{dx^2} + V(x)$ is said to be *PT* symmetric if the potential function satisfy $V(x) = V^*(-x)$. Such Hamiltonians possess a real eigen-spectrum. But if the imaginary component of $V(x)$ exceeds a certain threshold, the eigenspectrum ceases to be real resulting in spontaneous symmetry breaking [7].

In recent times, optics has proved to be a fertile ground for the investigation of *PT* symmetry both in linear as well as nonlinear systems. It was A. Ruschhaupt, F. Delgado and J. G. Muga [8], who first proposed the idea in 2005 in the context of planar slab waveguides. Moreover, the isomorphism of the paraxial equation of diffraction [8] with Schrodinger's wave equation presented a feasible way to explore *PT* symmetry in the field of optics provided one can appropriately synthesize the refractive index profile of the system to satisfy, $n(x) = n^*(-x)$. This analogy enabled researchers to observe the first experimental evidence of *PT* symmetry in optical waveguide structures [7]. Since then there has been numerous works on *PT* symmetry in optics, both in experimental as well as theoretical settings. PT Symmetry is studied in various contexts such as: Bragg solitons in nonlinear PT-symmetric periodic potential [9], continuous and discrete Schrodinger systems with PT-symmetric nonlinearities [10-12], bright and dark solitons and existence of optical rogue waves [13-19], modulation instability in nonlinear PT-symmetric structures [20-21], optical oligomers [22-29], optical mesh lattices [30-33], unidirectional invisibility [34], non-reciprocity and power oscillations [35,36], field propagation in linear and nonlinear stochastic *PT* coupler [37], optical mode conversion and transmission on photonic circuits [38] and so on.

In coupled waveguide systems, the *PT* phase transition is characterized by exponential growth and decay of optical power. Such systems have been studied in great detail [22]. The equations governing such systems can be analytically solved if the system is devoid of any form of nonlinearity. But in the presence of nonlinearity, analytical solution is not possible and prior assumptions are required. For instance, in Ref. [22], the system was studied taking stationary waves into consideration, whereas in Ref. [23], Stokes' parameters were used to study the conserved quantities. In the same line of research, this work aims to study the nonlinear *PT* symmetric dimer from a dynamical point of view. We have considered a waveguide coupler as our 'dimer' system. A thorough stability analysis of the fixed or stationary points in the system is carried out. This gives us a clearer and detailed interpretation of the dynamics subjected to different initial conditions. In our discussion, we will use the terms fixed points and stationary states interchangeably.

The article is organized as follows. In Section II, the theoretical model is described briefly. Section III presents and discusses the stability analysis of the ground state of the coupler below and above the *PT* threshold. It also discusses the non-zero stationary states of the configuration in the unbroken regime and at the phase transition point followed by conclusion in Section IV.

## 2. THE MODEL

The *PT* symmetric nonlinear coupler is a configuration consisting of two waveguides in close proximity so as to facilitate the transfer of optical power from one waveguide to the other via evanescent coupling. One waveguide can amplify the input optical signal and the other can attenuate the signal by the same proportion. The equations governing the dynamics of such a configuration are given by [23]:

$$i\frac{da_1}{dz} = i\gamma a_1 + Ca_2 + |a_1|^2 a_1$$
$$i\frac{da_2}{dz} = -i\gamma a_2 + Ca_1 + |a_2|^2 a_2 \qquad (1)$$

Here, $a_1$ and $a_2$ are the field amplitudes and $\gamma$ characterizes the gain/loss in the two channels and $C$ is the coupling constant. Both waveguides portray Kerr nonlinearity of equal strength.

In the absence of Kerr nonlinearity, the *PT* threshold is given by $\gamma_{th} = C$. But adding the nonlinearity changes the entire dynamics of the system. The reason is that once the system is modified with the inclusion of nonlinear terms, the initial conditions will play a major role in the dynamics of optical power evolution [28]. It must be noted here that the *PT* threshold of the linear coupler will be used as a reference point to study the stability analysis.

## 3. STABILITY ANALYSIS AND DISCUSSION

We first consider the ground state of the coupler defined by: $a_1 = a_2 = 0$. This set of initial conditions corresponds to unexcited waveguides. To ascertain the stability of the ground state, we expand the differential equations using the prescription $a_1 = x_1 + iy_1$ and $a_2 = x_2 + iy_2$. Eq. (1) can then be re-written as follows:

$$\dot{x}_1 = \gamma x_1 + Cy_2 + (x_1^2 + y_1^2)y_1 \qquad (2a)$$
$$\dot{y}_1 = \gamma y_1 - Cx_2 - (x_1^2 + y_1^2)x_1 \qquad (2b)$$
$$\dot{x}_2 = -\gamma x_2 + Cy_1 + (x_2^2 + y_2^2)y_2 \qquad (2c)$$
$$\dot{y}_2 = -\gamma y_2 - Cx_1 - (x_2^2 + y_2^2)x_2 \qquad (2d)$$

The linearization Jacobian is given by

$$J = \begin{bmatrix} \gamma + 2x_1 y_1 & x_1^2 + 3y_1^2 & 0 & C \\ -(3x_1^2 + y_1^2) & \gamma - 2x_1 y_1 & -C & 0 \\ 0 & C & -\gamma + 2x_2 y_2 & x_2^2 + 3y_2^2 \\ -C & 0 & -(3x_2^2 + y_2^2) & -\gamma - 2x_2 y_2 \end{bmatrix} \qquad (3)$$

The Jacobian eigenvalues are calculated to be $\lambda = \pm\sqrt{\gamma^2 - C^2}$. For $\gamma < C$, all eigenvalues of the Jacobian are purely imaginary indicating that the ground state is a non-hyperbolic fixed point [39]. Linear stability analysis fails if the fixed point under consideration is non-hyperbolic [40]. In mathematical terms, if all the eigenvalues are purely imaginary, the fixed point is classified as non-hyperbolic. In such a case, numerical solution of the system, under a suitably chosen perturbation, reveals the exact nature of the fixed point. On the other hand, if one or some of the eigenvalues contain non-zero real part the fixed point is categorized as hyperbolic. In such cases, linear stability analysis is sufficient. Above the *PT* threshold, the Jacobian has two positive and two negative eigenvalues. This means that the ground state is an unstable saddle fixed point and even the slightest excitation imparted to one of the waveguide will lead to an exponential growth and decay of the optical power in the two waveguides. But within the unbroken regime, an in-depth analysis of the system reveals that our model admits non-zero fixed points and in order to evaluate them, we resort to a polar form of Eq. (2) which would provide us with much more interesting picture of the non-zero fixed points and the dynamics therein.

Using $\dot{r}_i = \frac{x_i \dot{x}_i + y_i \dot{y}_i}{r_i}$ and $\dot{\theta}_i = \frac{x_i \dot{y}_i - y_i \dot{x}_i}{r_i^2}$, Eq. (2) is rewritten as follows:

$$\dot{r}_1 = \gamma r_1 + Cr_2 \sin(\theta_2 - \theta_1) \qquad (4a)$$
$$\dot{r}_2 = -\gamma r_2 - Cr_1 \sin(\theta_2 - \theta_1) \qquad (4b)$$
$$\dot{\theta}_1 = -C\frac{r_2}{r_1}\cos(\theta_2 - \theta_1) - r_1^2 \qquad (4c)$$
$$\dot{\theta}_2 = -C\frac{r_1}{r_2}\cos(\theta_2 - \theta_1) - r_2^2 \qquad (4d)$$

The equations 4(c) and 4(d) corresponding to the phases $\theta_1$ and $\theta_2$ can be coupled using $\theta = \theta_2 - \theta_1$ as follows:

$$\dot{\theta} = C\left(\frac{r_2}{r_1} - \frac{r_1}{r_2}\right)\cos(\theta) + (r_1^2 - r_2^2) \qquad (5)$$

We define $\theta$ as the relative phase lag parameter. The non-zero stationary states of our configuration are found to be:

$$r_1 = r_2 = r^* \qquad 6(a)$$
$$\theta = 2\pi - \sin^{-1}\left(\frac{\gamma}{C}\right). \qquad 6(b)$$

Physically, this means that the stationary states correspond to light inputs of equal optical power and the field in one waveguide lags behind by $\theta$ in phase with respect to the other. It can be clearly seen from Eq. 6(b) that the nonlinear dimer admits no fixed points for $\gamma > C$. This suggests that even in the presence of nonlinearity, the threshold point remains unchanged. In a linear coupler, the *PT* threshold dictates the evolutionary characteristics of optical power in the two waveguides. On the other hand, in a nonlinear coupler, the *PT* threshold governs the existence of fixed points.

To analyze the dynamics and stability of the non-zero fixed points, we will need to use the linearization Jacobian $J^*$ corresponding to Eqs. 4(a), 4(b) and (5), given as follows:

$$J^* = \begin{bmatrix} \gamma & C\sin(\theta) & Cr_2\cos(\theta) \\ -C\sin(\theta) & -\gamma & -Cr_1\cos(\theta) \\ \alpha & \beta & \delta \end{bmatrix} \quad (7)$$

Here, $\alpha = -C\left(\frac{r_1^2+r_2^2}{r_2 r_1^2}\right)\cos(\theta) + 2r_1$, $\beta = C\left(\frac{r_1^2+r_2^2}{r_1 r_2^2}\right)\cos(\theta) - 2r_2$ and $\delta = C\left(\frac{r_1^2-r_2^2}{r_1 r_2}\right)\sin(\theta)$.

The eigenvalues of $J^*$ are evaluated to be 0 and $\pm\sqrt{\gamma^2 - C^2(\sin\theta)^2 - 2ACr^*\cos\theta}$, where $A = \frac{2C\cos(\theta)}{r^*} - 2r^*$. The stability analysis will be now studied in two domains. We will study the stability of the non-zero stationary states (Eq. 6) for $\gamma < C$ followed by our analysis of the same at the threshold point $\gamma = C$.

For $\gamma < C$, the non-zero eigenvalues of $J^*$ can be further simplified to $\pm\sqrt{4C((r^*)^2 - C\cos(\theta))\cos\theta}$. From these eigenvalues, we can see that the initial launch conditions present another threshold on the stability of the system. This newfound threshold can be evaluated to be $r_{th}^* = \sqrt{C\cos\theta}$, which on further simplification gives

$$r_{th}^* = (C^2 - \gamma^2)^{\frac{1}{4}} \quad (8)$$

If $r^* > r_{th}^*$, one of the Jacobian eigenvalue is positive which means that the fixed point is an unstable saddle node. For instance, when $\gamma = 0.9$ and $C = 1$, we have $r_{th}^* = 0.6602$. Numerical solution of our configuration clearly shows that if the initial conditions are chosen such that $r^* < r_{th}^*$, the optical power in both waveguides exhibits aperiodic oscillations provided one of the launch conditions is subjected to some order of perturbation. This has been clearly illustrated in Fig. 1.

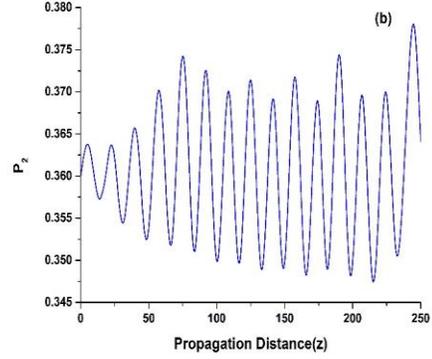
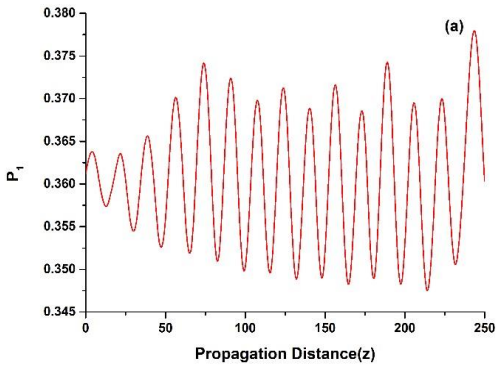
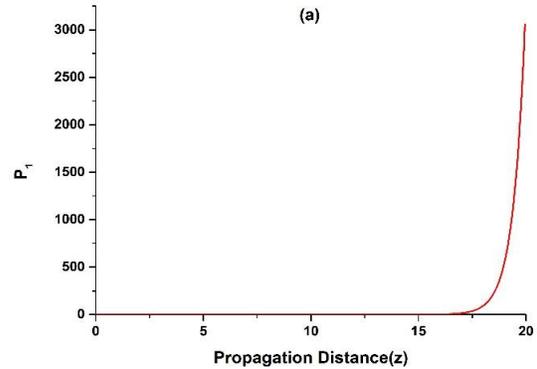

Fig.1. Spatio-power evolution in (a) waveguide 1 and (b) waveguide 2 for $\gamma = 0.9$. Initial conditions: $r_1 = 0.6 + 10^{-4}$, $r_2 = 0.6$ and $\theta = 5.1634$.

In Fig.1, the initial launch conditions for both waveguides have been chosen below the threshold (Eq. 8) and one of them have been subjected to a perturbation of the order of $10^{-4}$. This has been done in order to visualize the stability of the initial launch conditions under the influence of fluctuations. The gain/loss coefficient has been set at $\gamma = 0.9$ and accordingly, the threshold is evaluated to be: $r_{th}^* = 0.6602$. So, in order to satisfy $r^* < r_{th}^*$, we chose $r_1 = 0.6 + 10^{-4}$ and $r_2 = 0.6$. And the relative phase lag $\theta$ has been decided in accordance with Eq.6 (b). This gives rise to aperiodic oscillations in the spatial propagation of optical power in their respective waveguides. On a closer inspection, we can clearly see that optical power fluctuates on the order of $10^{-2}$ in both waveguides.

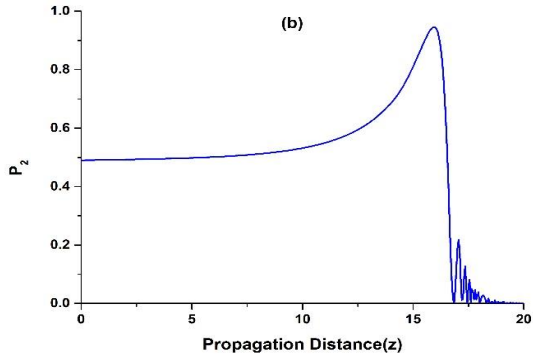

Fig. 2. Spatio-power evolution in (a) waveguide 1 and (b) waveguide 1 for $\gamma = 0.9$. Initial Conditions: $r_1 = 0.7 + 10^{-4}, r_2 = 0.7$ and $\theta = 5.1634$.

On the other hand, if $r^* > r_{th}^*$, power evolution in both waveguides will exhibit characteristics similar to that of a linear coupler in the broken $PT$ regime. This has been shown in Fig.3. The initial launch conditions are chosen in a similar manner as considered for the case $r^* < r_{th}^*$. Fig. 2 depicts the exponential growth and decay of optical power in the two channels. As it has been pointed out before, this happens because one of the eigenvalues of $J^*$ is real positive and our stability analysis clearly shows that the fixed point is an unstable node. Similar dynamics in spatial propagation of optical power has been observed in the linear coupler when the gain/loss coefficient is taken above the $PT$ threshold. But in our analysis, we have seen that even the choice of initial conditions matters a lot and this can be attributed to the presence of the nonlinearity in our model. To present this aspect of our system in a clearer manner, Fig. 3 depicts the contour plot of the real part of the Jacobian eigenvalues. Moreover, we have also included the plot of Eq. 8 (blue colored line). This has been done in order to dwell on the lower bound of $r$ above which the Jacobian has real eigenvalues. Above this line, the system is in an unstable state for any given initial launch conditions.

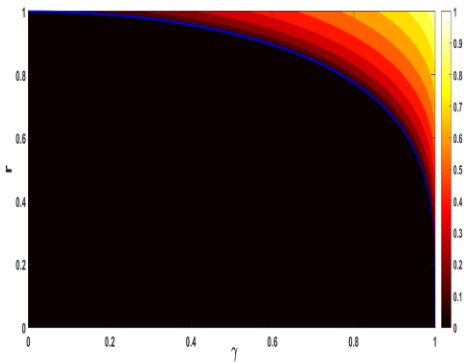

Fig. 3. Contour plot of the real component of the Jacobian eigenvalues versus $r$ and $\gamma$. The blue colored line is the plot of Eq. 8.

For the analysis with regards to the dynamics of the system at the $PT$ threshold, we choose $C = 1$. At the $PT$ threshold, the relative phase lag, given by Eq. 6(b), is $\theta = 3\pi/2$. It is very interesting to note here that under these circumstances, all eigenvalues of the linearization Jacobian $J^*$ are zero. A study on the numerical solution and the corresponding phase plane analysis has been done by choosing a set of initial conditions corresponding to the fixed points at the $PT$ threshold. Since, all the Jacobian eigenvalues are zero at the $PT$ threshold, we have the freedom to choose $r^*$ at any arbitrary value, whereas the relative phase lag is set at $\theta = 3\pi/2$. We present two aspects of our configuration, one of which is the attractor aspect and the other is the chaotic behavior in the real and imaginary component of the field amplitudes. The gain/loss coefficient is set at $\gamma = 0.95$.

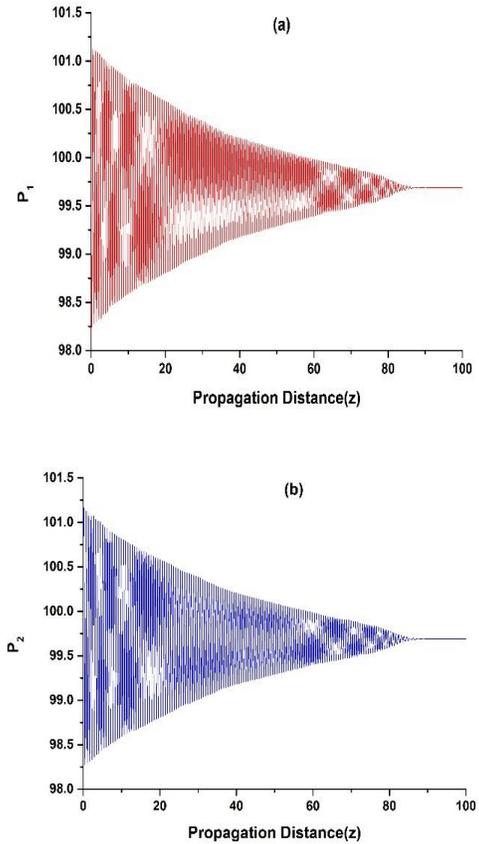

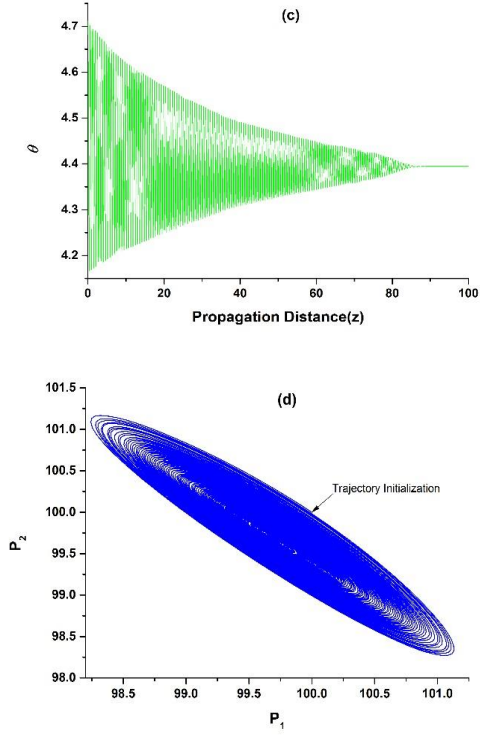

Fig. 4. (a) Spatial evolution of optical power in waveguide 1 and (b) waveguide 2 (c) Spatio-evolution of relative phase lag $\theta$ and (d) Phase plane of optical power in both waveguides. Initial Conditions: $r_1 = r_2 = 10$ and $\theta = 3\pi/2$.

Fig.4 (a-c) depicts decay in oscillations of the evolution of optical power and the relative phase lag along the propagation distance. It must be noted here that this happens when the initial conditions are chosen for $\gamma = C$ using Eq. 6(a-b) and a slight change in the gain/loss coefficient is introduced. We can see that the oscillations decay in the initial stage and slowly approaches a constant value. Moreover phase plane analysis (Fig. 4(d)) of the optical power in both waveguides shows an inspiral trajectory and this is sufficient for us to ascertain the existence of an attractor in our configuration. Numerical computation shows that the attractor point is located at $r_1 = r_2 = 9.9844$ and $\theta = 4.394852$. From Eq. 6(b), the attractor point concerning the relative phase lag $\theta$ corresponds to $\gamma = 0.95$. Hence, we can infer that if our initial conditions are chosen in accordance with the fixed points at the $PT$ threshold, any disturbance in $\gamma$ will redirect the trajectory to an attractor point. This attractor point coincides with the fixed point of the new $\gamma$. Moreover, we can see that there is some loss in the initial optical power after an initial run of decaying oscillations. From this we can infer that the loss in optical power is compensated by a change in the relative phase lag between the two optical fields.

On the other hand, it must be noted here that even though the relative phase lag $\theta$ becomes constant after the field amplitudes has propagated some distance, the individual phases of the field amplitudes $\theta_1$ and $\theta_2$ do not stabilize. The reason behind this can be attributed to the fact that the fixed points we have been dealing with so far are those of Eqs. 4(a), 4(b) and (5). So to extract the information on the evolution of $\theta_1$ and $\theta_2$, we need to proceed with the numerical solution of Eqs. 4(a-d). This way, we can demonstrate the dynamics of the real and imaginary component of the field amplitudes.

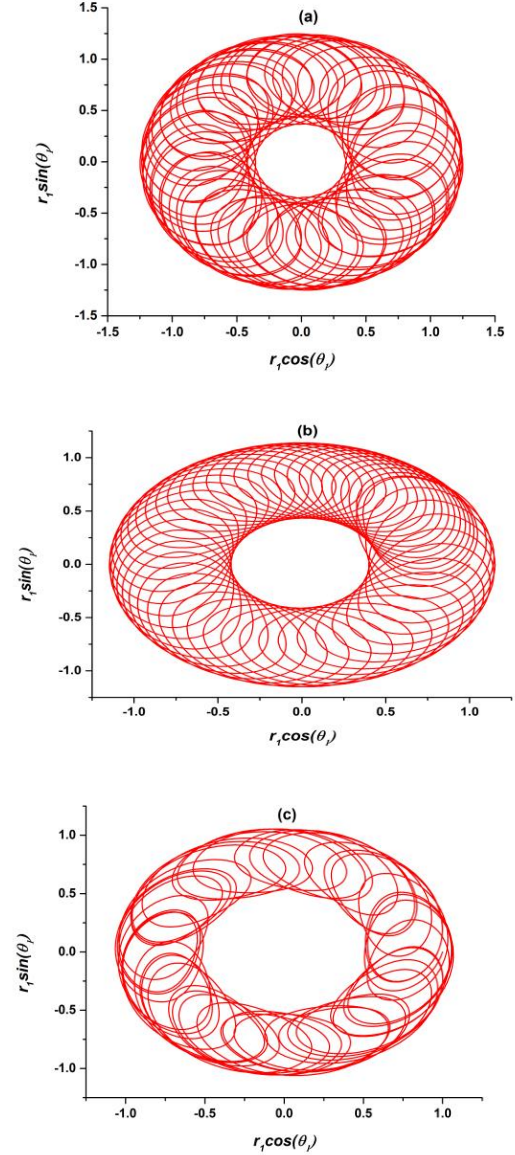

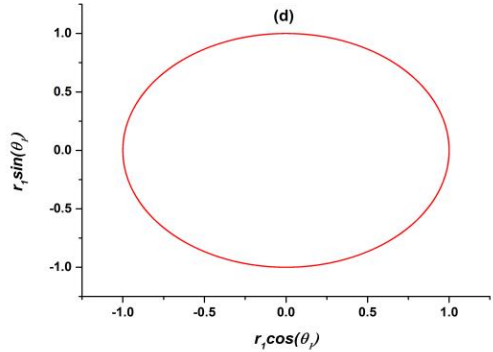

Fig. 5. Phase Plane trajectory of the real and imaginary component of the field amplitude in waveguide 1 for (a) $\gamma = 0.25$, (b) $\gamma = 0.5$, (c) $\gamma = 0.75$ and (d) $\gamma = 1$. Initial Conditions: $r_1 = r_2 = 10$, $\theta_1 = 0$ and $\theta_2 = 4.7124$.

Phase plane analysis of the real and imaginary component of the field amplitude (Fig. 5) in the first waveguide shows a toroidal trajectory, which vanishes as $\gamma$ approaches the *PT* threshold. In Fig. 5(a-c), we can see the presence of two circular orbits within which the real and imaginary components oscillate. As $\gamma$ increases, the two orbits approach closer and closer and finally they merge with each other. On a closer look, we can see that in Fig. 5(a), there is a third circular orbit, which lies very close to the inner orbit. But it disappears in Fig. 5(b). In Fig. 5(c), radius of the inner circular orbit further increases and in Fig. 5(d), it merges with that of the outer circular orbit when $\gamma = C$. This is similar to Sil'nikov orbits [41,42]. The presence of such orbits implies a chaotic trajectory, which vanishes under the influence of some parametric changes. In our case, the chaotic trajectory disappears at the *PT* threshold and it takes the form of a limit cycle.

## 4. CONCLUSIONS

In conclusion, we have investigated several aspects of the nonlinear *PT* symmetric coupler from a dynamical perspective. As opposed to linear PT-coupler where the *PT* threshold dictates the evolutionary characteristics of optical power in the two waveguides, in a nonlinear coupler, the *PT* threshold governs the existence of stationary points. We have found that the stability of the ground state undergoes a phase transition when the gain/loss coefficient is increased from zero to beyond the *PT* threshold. Moreover, in the unbroken regime, we find that the instabilities in the initial launch conditions can trigger an exponential growth and decay of optical power in the waveguides. Also, it can redirect the spatial power evolution into aperiodic oscillations. The attractor behavior of the system has also been studied under changes in the gain/loss coefficient. From our phase plane analysis, we can ascertain that such a system exhibits self-stabilizing characteristics. And finally, we have shown that the phase plane trajectory of the real and imaginary component of the field amplitudes is a toroidal chaotic trajectory. Such chaotic trajectory could be controlled with judicious choice of waveguide parameters.

## ACKNOWLEDGEMENTS

J.P.D. and A.K.S. would like to acknowledge the financial support from DST, Government of India (Grant No. SB/FTP/PS-047/2013).